\documentclass[conference]{IEEEtran}
\usepackage{cite}
\usepackage{amsmath,amssymb,amsfonts}
\usepackage{algorithmic}
\usepackage{graphicx}
\usepackage{textcomp}
\usepackage{xcolor}
\usepackage{comment}
\usepackage{listings}
\usepackage{fancyvrb}

\usepackage{blkarray}
\usepackage{amsmath}
\usepackage{dsfont}
\usepackage{amsthm}
\usepackage{lipsum}
\usepackage{graphicx}

\usepackage[utf8]{inputenc}
\usepackage[english]{babel}

\usepackage[T1]{fontenc}
\usepackage{fancyhdr}

\def\BibTeX{{\rm B\kern-.05em{\sc i\kern-.025em b}\kern-.08em
    T\kern-.1667em\lower.7ex\hbox{E}\kern-.125emX}}
\begin{document}

\title{
\begin{LARGE}{Agent Based Virus Model using NetLogo: Infection Propagation, Precaution, Recovery, Multi-site Mobility and (Un)Lockdown}\end{LARGE}
}

\author{\IEEEauthorblockN{Dibakar Das}
\IEEEauthorblockA{
{IIIT Bangalore}, India \\
dibakard@ieee.org}
}
\maketitle

\begin{abstract}
This paper presents a novel virus propagation model using NetLogo. The model allows agents to move across multiple sites using different routes. Routes can be configured, enabled for mobility and (un)locked down independently. Similarly, locations can also be (un)locked down independently. Agents can get infected, propagate their infections to others, can take precautions against infection and also subsequently recover from infection. This model contains certain features that are not present in existing models. The model may be used for educational and research purposes, and the code is made available as open source. This model may also provide a broader framework for more detailed simulations. The results presented are only to demonstrate the model functionalities and do not serve any other purpose.

\end{abstract}


\section{Introduction}
Agent based models have been explored for virus spread for a considerable period of time. These models have evolved over time based on the newer requirements. NetLogo \cite{cite_netlogo_website} is one of the agent based modeling platform that has been used to model virus spread as well.

Several agent based models using NetLogo have been proposed in the literature. NetLogo model library contains a basic model of infection spread \cite{cite_virus_default_netlogo}. 
\cite{cite_covid19_extended_default_netlogo} extends \cite{cite_virus_default_netlogo} with population density, degree of immunity, infectiousness, duration of infection. \cite{cite_covid19_solo_epDIM_netlogo} presents a NetLogo model to include travel, isolation, quarantine, inoculation, links between individuals, and spread of an infectious disease in a closed population. \cite{cite_covid19_disease_spread_probability_netlogo} presents a probabilistic model of COVID-19 spread using NetLogo. 
All these previous works focus on different aspects of virus spread. This paper presents a novel NetLogo based virus infection propagation considering multi-site mobility, route and site (un)lockdowns, precautions and recovery. To the best of the knowledge of the author, none of the previous works addresses these aspects.

The paper is organized as follows. The proposed model is described in section \ref{section_model}. Implementation aspects are explained in \ref{section_implementation}. Results to demonstrate the model are presented in section \ref{section_results}. Conclusions are drawn in section \ref{section_conclusion}.
\section{Model Description}\label{section_model}
\begin{figure}[ht]
\centering
\includegraphics[width=\columnwidth]{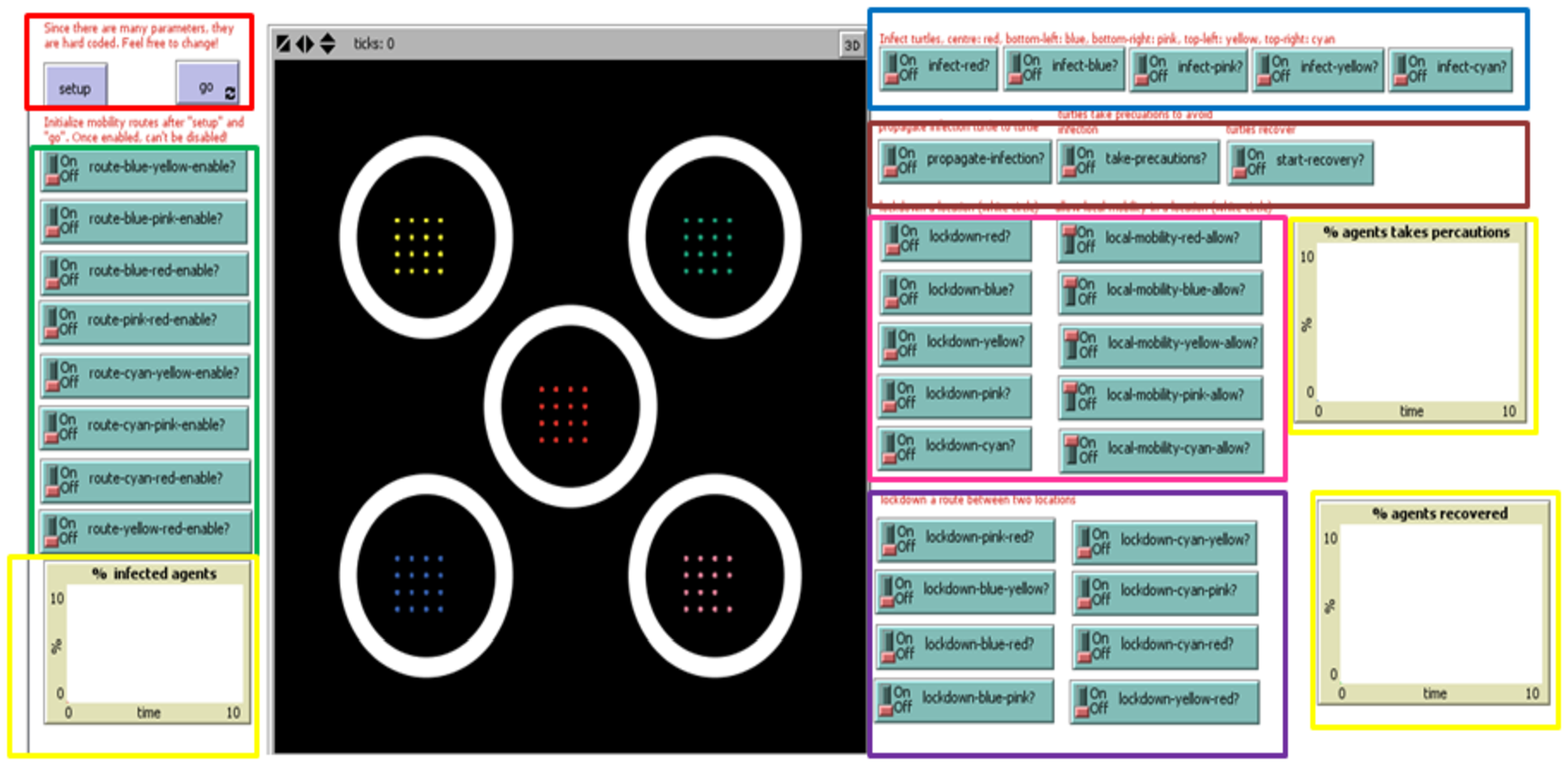}
\caption{Complete GUI options}
\label{fig_full_gui_inkscape}
\end{figure}

Fig. \ref{fig_full_gui_inkscape} shows the features available in the proposed model. There are 5 locations and 8 routes in the model. Since, there are many parameters in the model, some of them are configurable in the GUI and several others are hard coded to avoid GUI to be too much cluttered. The model runs based on the GUI configuration and the hard coded parameters. The hard coded parameters may be modified to suit the needs of simulation. Red box in Fig. \ref{fig_full_gui_inkscape} contains the  \verb|setup| and  \verb|go| procedures. The green box on the left shows the routes that can be configured with switch options. Once a route is configured, it cannot be disabled till next run. The blue box on top right of Fig. \ref{fig_full_gui_inkscape} contains 5 switch options to infect turtles (agents) in the 5 locations (white circles), namely, red (centre), blue (bottom-left), pink (bottom-right), cyan (top-right) and yellow (top-left). The brown box below the blue contains 3 switch options. One of them is to propagate infection to other turtles. The next switch option is to enable precautions for the turtles. The last switch option is for turtle recovery. The pink box on the right contains 2 columns, each with 5 switch options. Each row of 2 switch options allows a location (white circles) to be locked down and local mobility to be disabled. If a location is locked down, mobility of turtles along its configured routes are disabled. If local mobility is also disabled along with location lock down, then turtles freeze their movements in that location. To unlock a location (white circle), set the corresponding switch to \verb|Off| position, enable local mobility (if disabled) and then enable mobility of turtles explicitly along the associated configured routes. Each route can be individually locked down (and unlocked again) with the 8 switch options shown in the purple box shown on the right side of Fig. \ref{fig_full_gui_inkscape} below the pink box. There are 3 plots (yellow boxes) showing the percentage of turtles infected (bottom-left in Fig. \ref{fig_full_gui_inkscape}), percentage of recovered turtles (bottom-right) and percentage of turtles taking precautions (on top of the previous box). Turtles are circles by default, when they are infected their shape change to triangles, when they take precautions their shape are squares and when they recover their shapes are stars.

\begin{figure}[ht]
\centering
\includegraphics[width=\columnwidth]{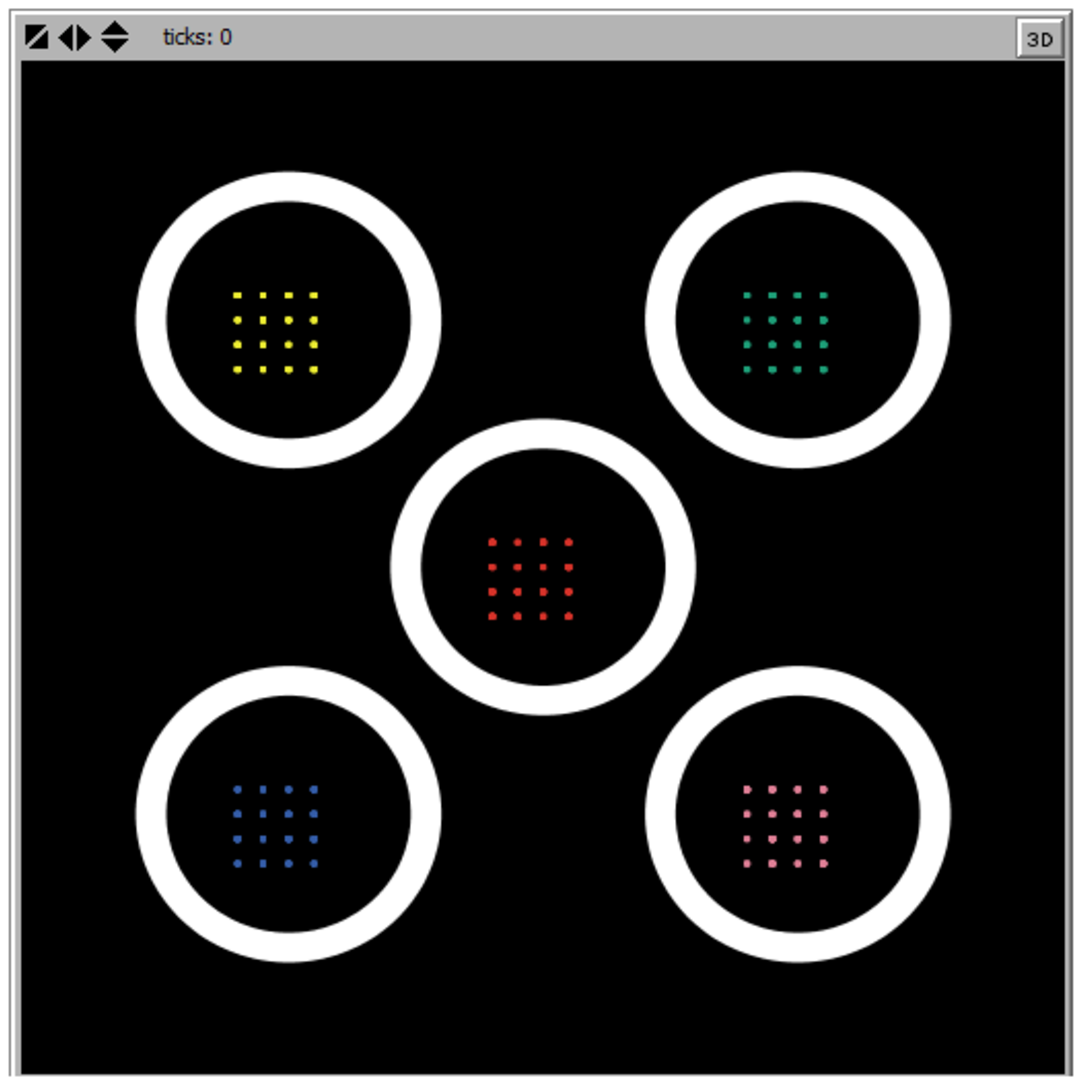}
\caption{State of turtles after ``setup''}
\label{fig_setup_inkscape}
\end{figure}

\begin{figure}[ht]
\centering
\includegraphics[width=\columnwidth]{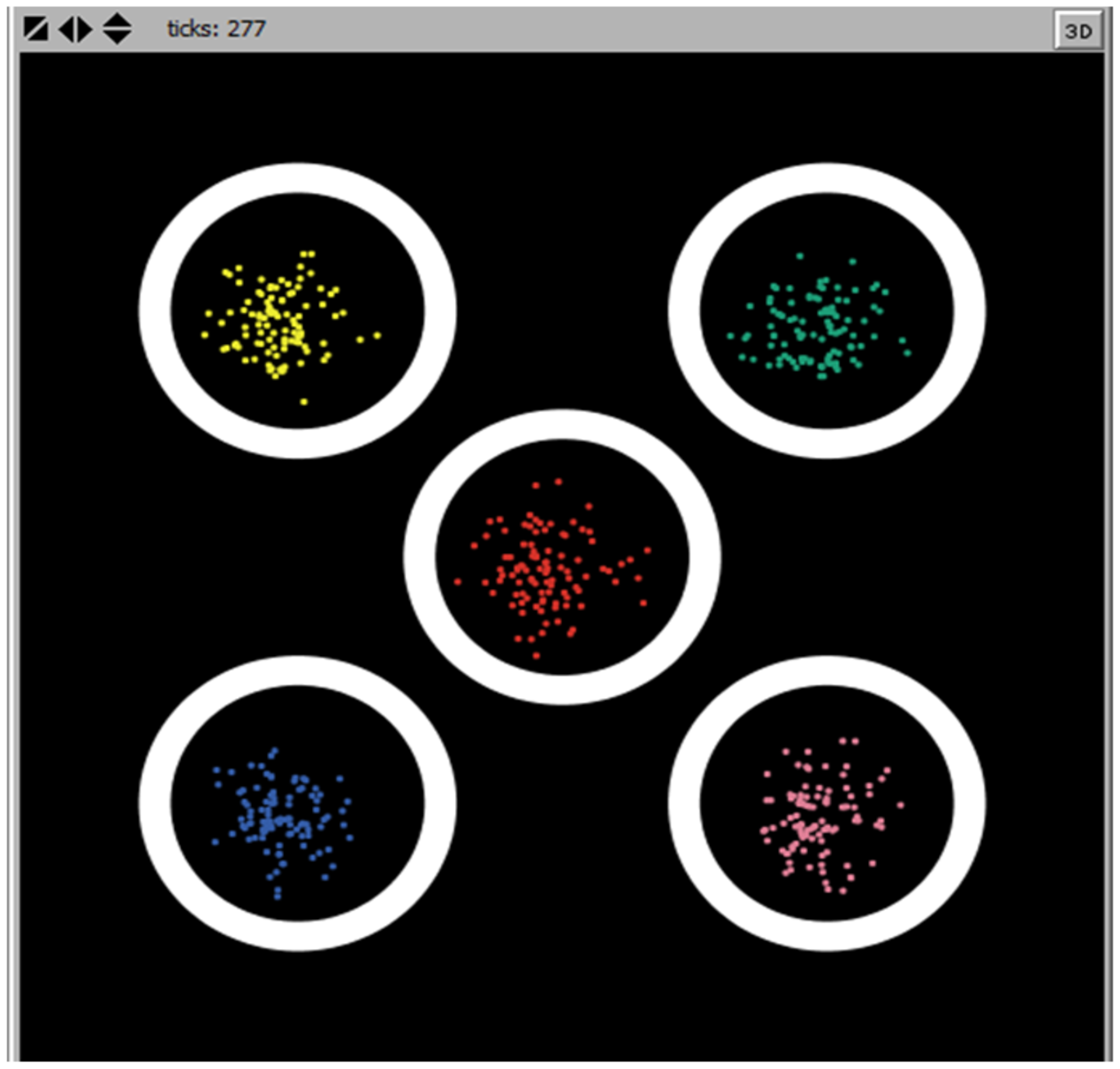}
\caption{State of turtles after ``go''}
\label{fig_go_inkscape}
\end{figure}

Fig. \ref{fig_setup_inkscape} shows the state of the turtles after pressing \verb|setup|. It has the 5 static white circles which are locations containing the respective turtles. These locations are called red, blue, pink, cyan and yellow locations based on the turtles' colour.
Fig. \ref{fig_go_inkscape} shows the state of the turtles after pressing \verb|go|. All the turtles move randomly in their respective 5 home locations.

\begin{figure}[ht]
\centering
\includegraphics[width=\columnwidth]{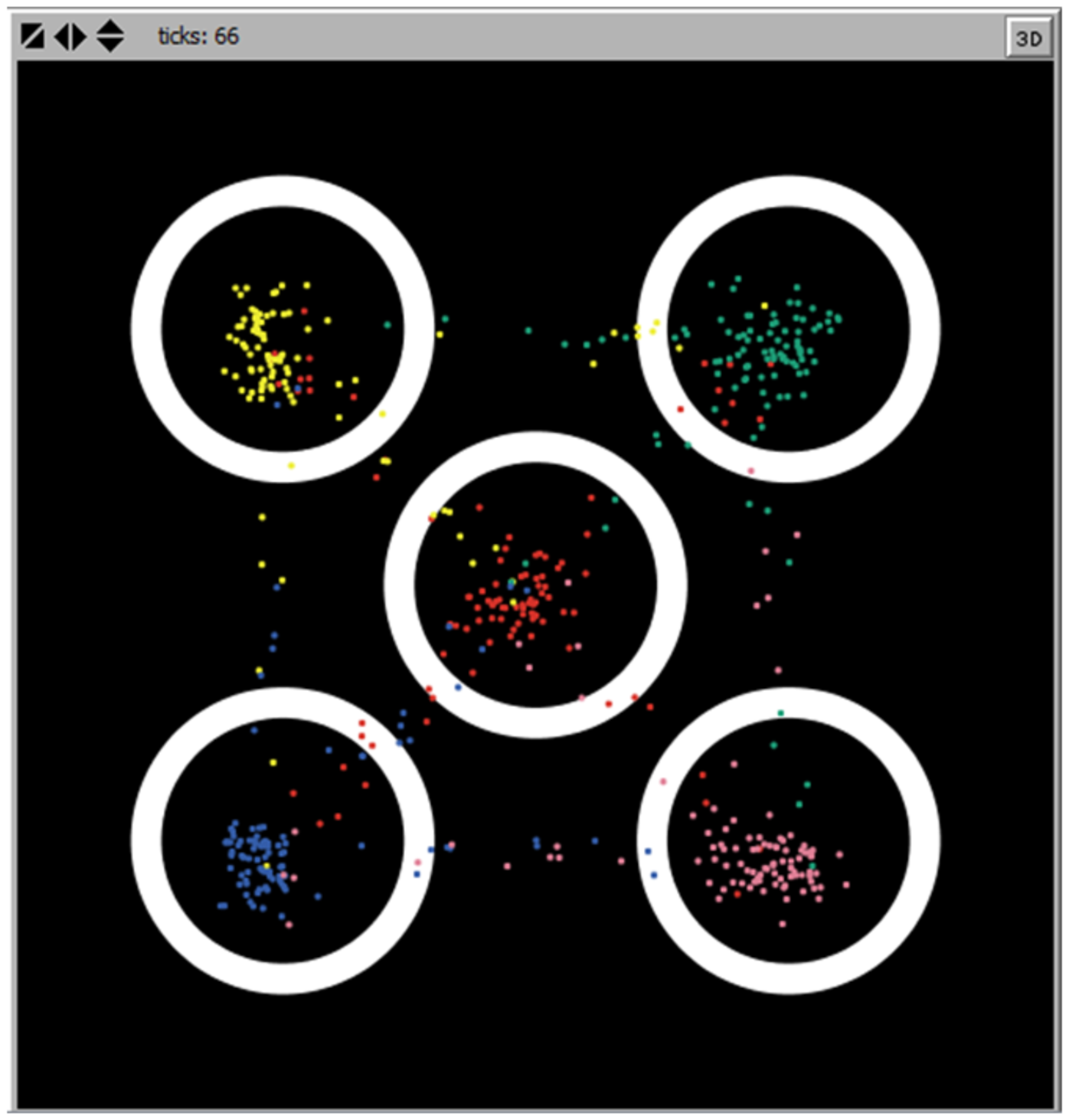}
\caption{State of the turtles when all the routes are configured}
\label{fig_route_enable_inkscape}
\end{figure}

Fig. \ref{fig_route_enable_inkscape} shows the state of the turtles when all the 8 routes are configured (using the switches in the green box on the left in Fig. \ref{fig_full_gui_inkscape}). Note that all the routes need not be configured. However, a route once configured cannot be disabled till the next run. Configured routes can be locked down and unlocked. For example, when \verb|route-blue-yellow-enable?| is switched \verb|On| some of blue turtles destined for location yellow (top-left white circle) move towards their destination. Similarly, some yellow turtles start moving towards blue location (bottom-left white circle).

\begin{figure}[ht]
\centering
\includegraphics[width=\columnwidth]{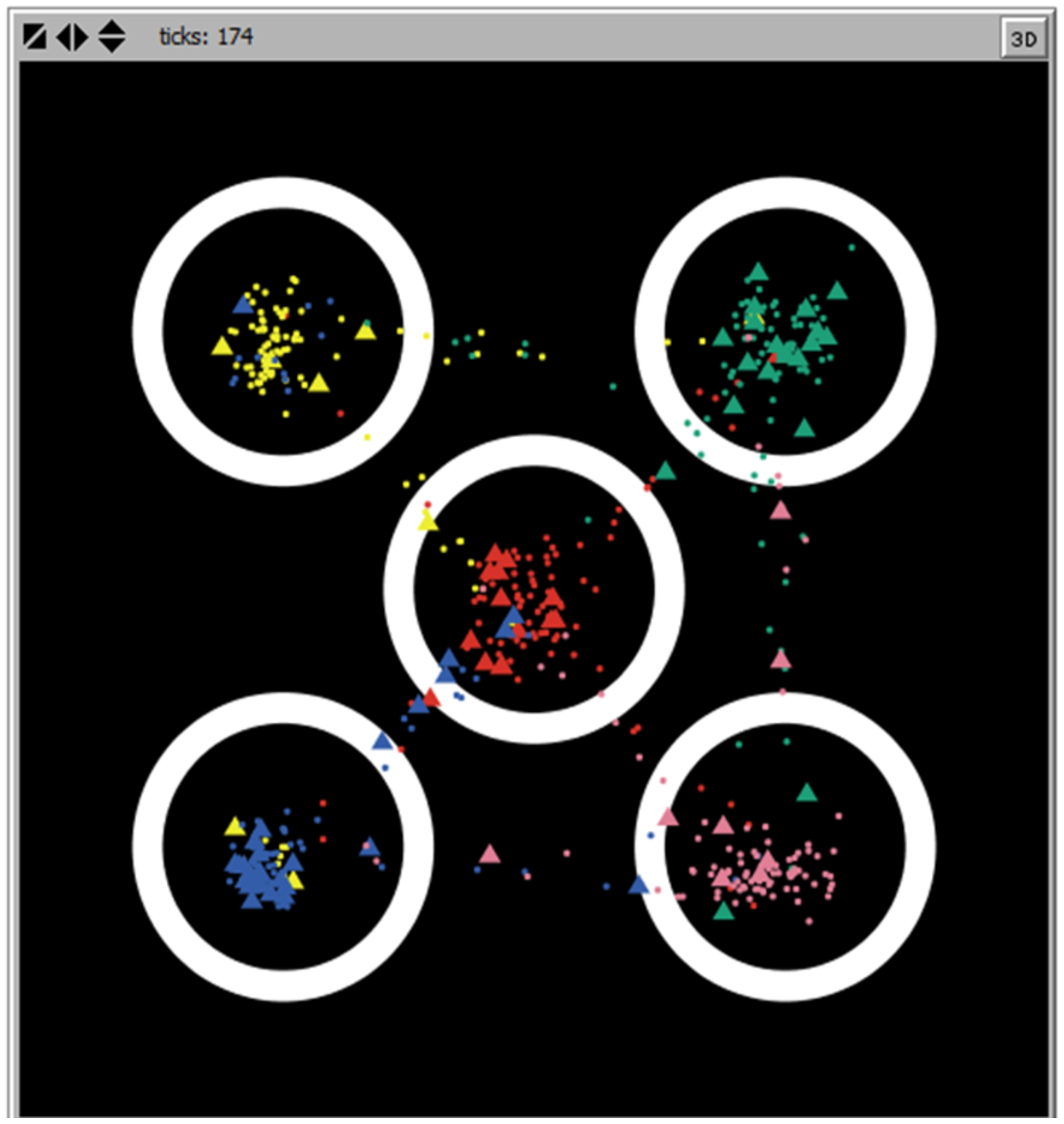}
\caption{State of the turtles when blue-yellow interface is locked down}
\label{fig_lockdown_interface_inkscape}
\end{figure}

Each of the 8 routes can be locked down independently (using the switched in purple box on the right in Fig. \ref{fig_full_gui_inkscape}). Fig. \ref{fig_lockdown_interface_inkscape} shows the state of the turtles when the blue-yellow route is locked down by setting \verb|lockdown-blue-yellow?| switch to \verb|On| state. Changing the switch to \verb|Off| state restores the mobility along that route provided either of the locations blue (bottom-left) and yellow (top-left) is not locked down (explained below).

\begin{figure}[ht]
\centering
\includegraphics[width=\columnwidth]{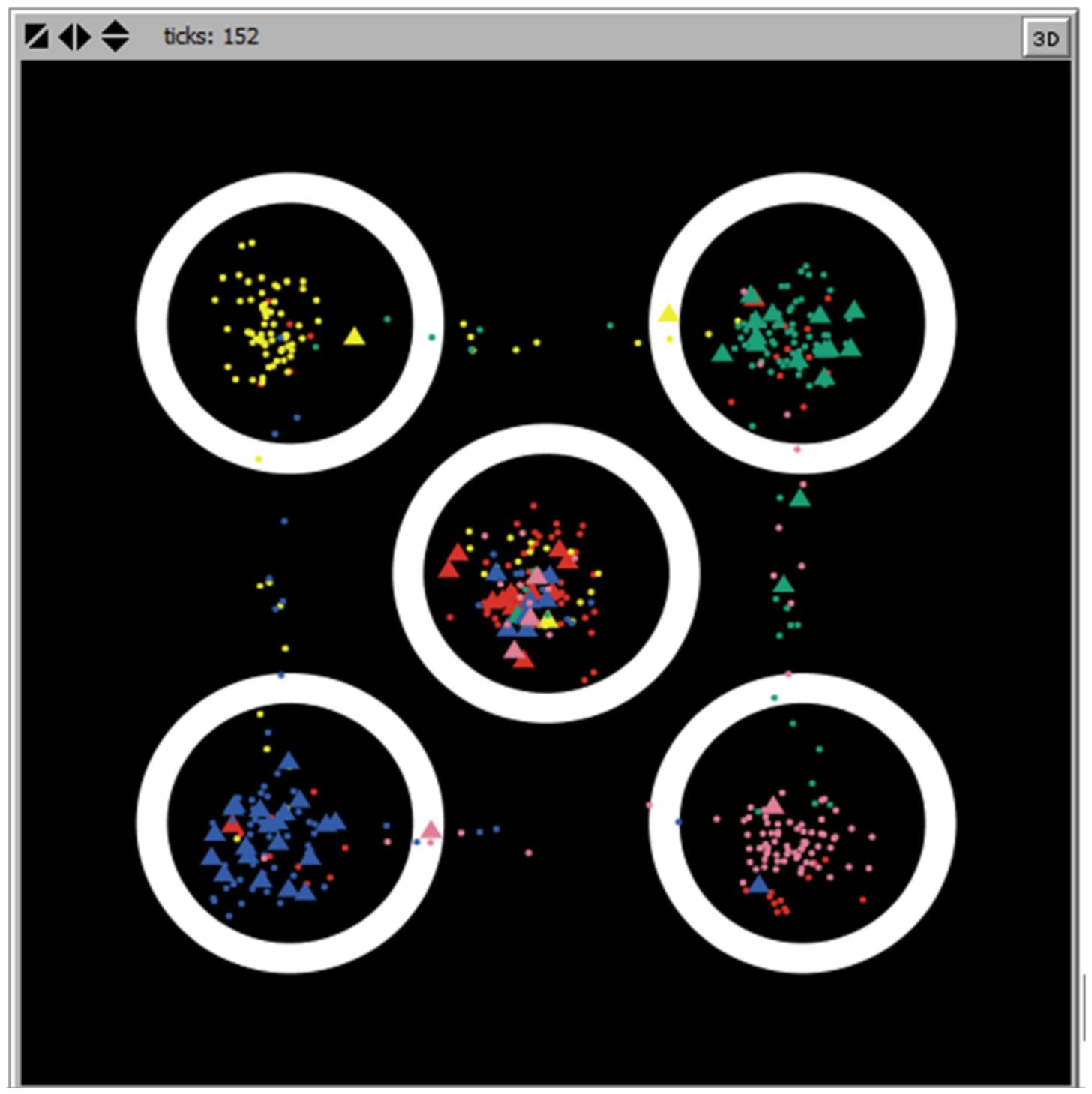}
\caption{State of the turtles when red location (centre) is locked down}
\label{fig_lockdown_location_inkscape}
\end{figure}

Each of the 5 locations can be locked down independently (pink box on the right in Fig. \ref{fig_full_gui_inkscape}). Locking down one of the location shuts down its associated routes. The turtles in the locked down location move within the region until the local mobility is also switched off. Fig. \ref{fig_lockdown_location_inkscape} shows the state of the turtles when the red location (centre) is locked down by setting \verb|lockdown-red?| to \verb|On| state. All its 4 routes to other locations are also locked down automatically, i.e., \verb|lockdown-pink-red?|, \verb|lockdown-blue-red?|, \verb|lockdown-cyan-red?| and \verb|lockdown-yellow-red?| are all set to \verb|On| states. If local mobility is also disabled by setting \verb|local-mobility-red-allow?| to \verb|Off| then the turtles is red location stop moving. To unlock red location, \verb|lockdown-red?| should be set to \verb|Off| state, \verb|local-mobility-red-allow?| set to \verb|On| state, and the 4 routes \verb|lockdown-pink-red?|, \verb|lockdown-blue-red?|, \verb|lockdown-cyan-red?| and \verb|lockdown-yellow-red?| should be set to \verb|Off| states.

\begin{figure}[ht]
\centering
\includegraphics[width=\columnwidth]{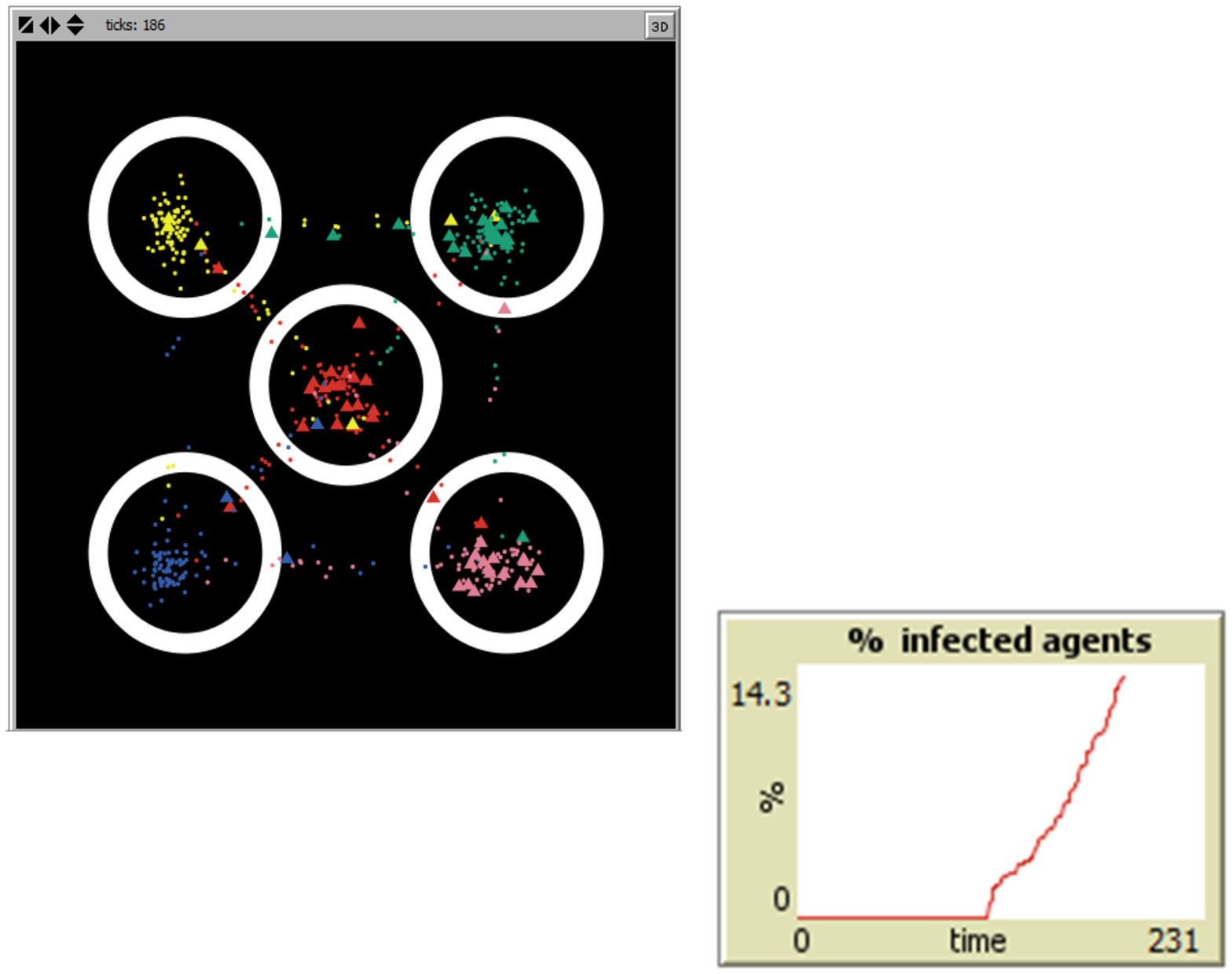}
\caption{Infection Propagation}
\label{fig_infect_propagate_inkscape}
\end{figure}

Turtles in each of the 5 locations can get infected independently by setting the switches \verb|infect-red?|, \verb|infect-blue?|, \verb|infect-pink?|, \verb|infect-cyan?| and \verb|infect-yellow?| to \verb|On| states (using the switches in the blue box on the right in Fig. \ref{fig_full_gui_inkscape}). Some of the randomly selected turtles in those locations get infected and change their shape from circles to triangles as shown in Fig. \ref{fig_infect_propagate_inkscape}. An infected turtle can infect another uninfected turtle with certain probability when the switch \verb|propagate-infection?| (switch in the brown box on the right in Fig. \ref{fig_full_gui_inkscape}) is set to \verb|On| state. The plot associated with Fig. \ref{fig_infect_propagate_inkscape} shows the percentage of turtles infected.

\begin{figure}[ht]
\centering
\includegraphics[width=\columnwidth]{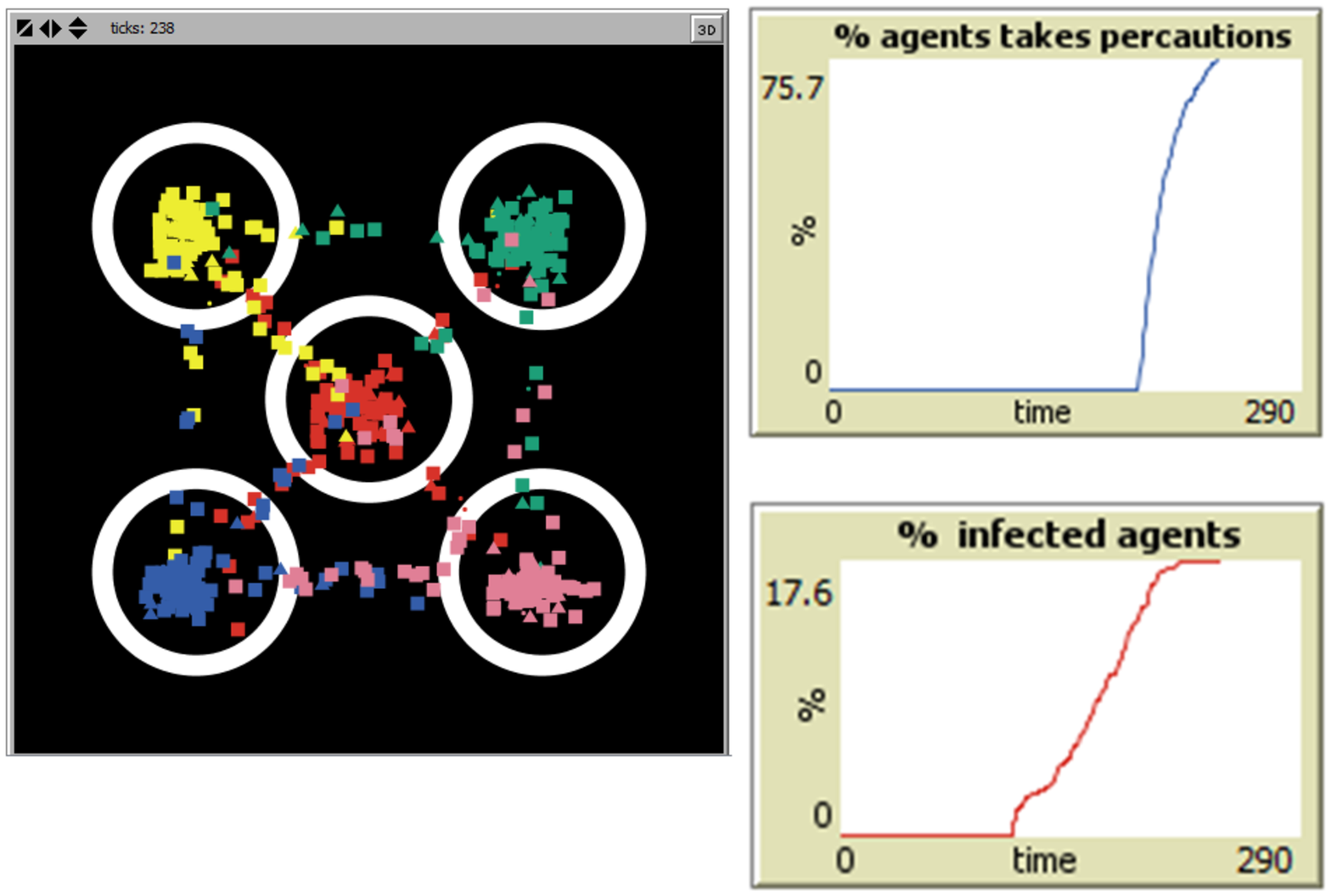}
\caption{Infection propagation and precaution}
\label{fig_precaution_inkscape}
\end{figure}

Turtles (randomly selected) start taking precautions when \verb|take-precautions?| is set to \verb|On| state (switch on the brown box on the right in Fig. \ref{fig_full_gui_inkscape}). Those turtles change their shapes to squares and they do not get infected by other infected turtles as shown in Fig. \ref{fig_precaution_inkscape}. The associated plot with this figure shows the percentage of turtles taking cautions. When more turtle take precautions, infection rate flattens.

\begin{figure}[ht]
\centering
\includegraphics[width=\columnwidth]{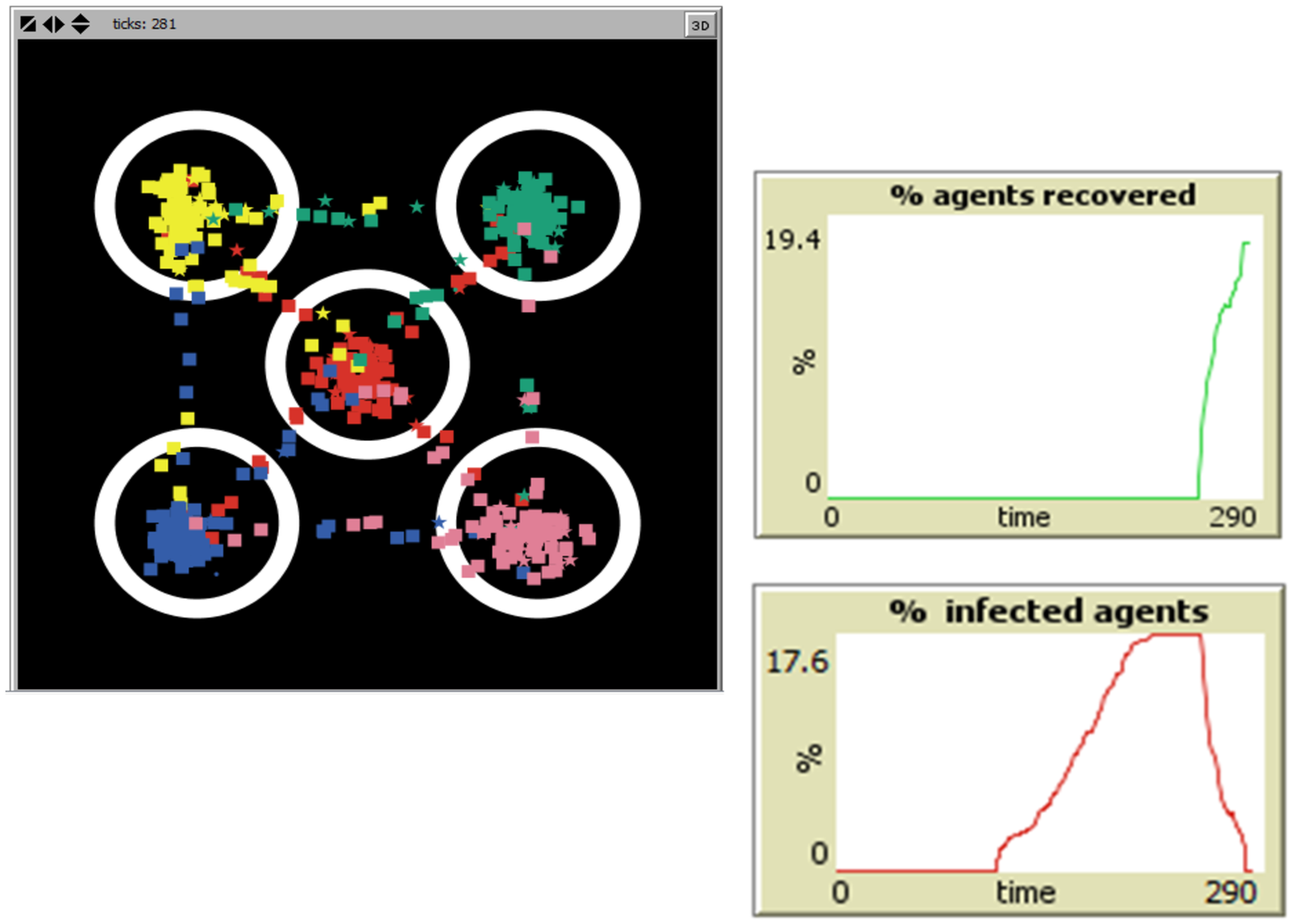}
\caption{Infection propagation and recovery}
\label{fig_recovery_inkscape}
\end{figure}

Once the \verb|start-recovery?| is set to \verb|On| (switch in the brown box on the right in Fig. \ref{fig_full_gui_inkscape}), infected turtles start recovery and change their shape to star as shown in Fig. \ref{fig_recovery_inkscape}. Associated plot in the figure shows that percentage of turtles recovered. Once the recovery process starts percentage of infected turtles begins to drop.

\section{Implementation}\label{section_implementation}
As the title suggests, the model is implemented in NetLogo \cite{cite_netlogo_website}. The code used minimal features of NetLogo to implement the model. The source code is open sourced \cite{cite_netlogo_my_model}. The implementation is simple to understand by going through the comments.  Since, there many parameters to control the model, several of them are hard coded to reduce the number of input options and keep the GUI simple.

The \verb|setup| function does a few typical NetLogo initialization steps and then calls \verb|static-setup| which does the initial configuration. Function \verb|static-setup| sets up the 5 locations, creates 100 turtles each of 5 different colours, and initializes private variables of turtles (own variables) e.g., their mobility destinations, etc., (\verb|my-heading|, \verb|my-immunity|, \verb|my-precaution?|, \verb|my-infection?|), and global boolean flags used by the model. Function \verb|static-go| called from \verb|go| moves the turtles within their location till local mobility is allowed in the respective places. Once routes (maximum of 8) are configured and not locked down, \verb|go| also enables mobility of some turtles along their designated routes to their respective designations, leaving rest of them in their homes. Function \verb|static-go| also calculates the percentage of turtles infected which is used for plotting. Function \verb|do-recovery| randomly selects a certain percentage (hardcoded) of infected turtles (shaped triangle) and moves them to recovered state (shaped star) when \verb|start-recovery?| is set to \verb|On| state. This function also calculates the percentage of recovered turtles used for plotting. Function \verb|do-precautions| calculates the number of turtles to be moved to precaution state, they don't get infected and their shape changes to squares when \verb|take-precautions?| is set to \verb|On| state. This function also calculates the percentage of turtles which take precautions not to get infected. Function \verb|start-infection| starts turtle infection based on the GUI switch options for the 5 locations. Once, a location is infected by setting the corresponding switch to \verb|On| state, it cannot be disabled. For each location, random number of turtles are set as infected and their shapes change to triangle. However, they start propagating to other neighbour turtles (with certain probability, based on \verb|my-immunity|, \verb|my-precaution?|) in function \verb|start-infection| when \verb|propagate-infection?| is set to \verb|On| state. This function also calculates the percentage of infected turtles used for plotting. Function \verb|static-lockdown-interface| is used to lock down any of the configured 8 routes independently. Once, a route is locked the turtles, destined to their respective location, defined by \verb|my-heading|, keep moving till they reach those locations. Further mobility of turtles stops thereafter. Function \verb|static-lockdown| locks down any of the 5 locations and their associated routes independently. Function \verb|toss-a-coin| generate some amount of randomness whether a turtle can get infected. Function \verb|mobility-direction| randomly allocates mobility direction to the turtles by setting the variable \verb|my-heading|.

Since, the model has many parameters, some of them has hardcoded. The most important ones are listed below.
\begin{itemize}
\item Number of turtles per location (fixed during initialization)
\item Number of turtles that move locally within a site (selected at random from a fixed set)
\item Number of turtles that move along the routes across location (selected at random from a fixed set)
\item Number of turtles that recovers (selected at random from a fixed set)
\item Number of turtles that take precautions (selected at random from a fixed set)
\item Number of turtles that get infected (selected at random from a fixed set)
\item Direction of movement of turtles that move across sites (selected at random from a fixed set)
\end{itemize}

\section{Results}\label{section_results}
The results presented in this section are only to demonstrate the model functionalities and do not serve any other purpose.

\begin{figure}[ht]
\centering
\includegraphics[width=\columnwidth]{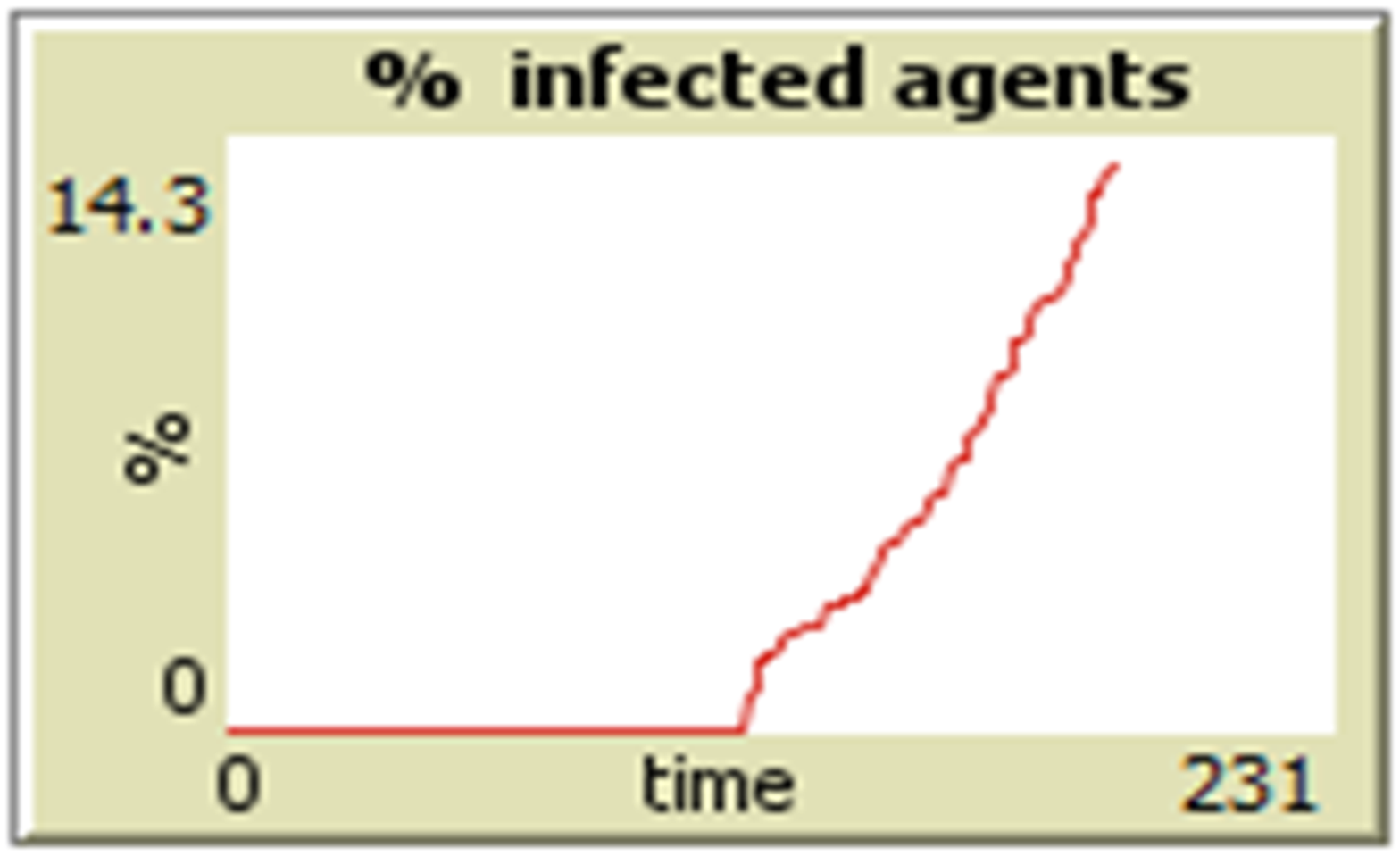}
\caption{Infection growth curve}
\label{fig_infection_plot_inkscape}
\end{figure}

The plot in Fig. \ref{fig_infection_plot_inkscape} shows how the infection grows with time.
\begin{figure}[ht]
\centering
\includegraphics[width=\columnwidth]{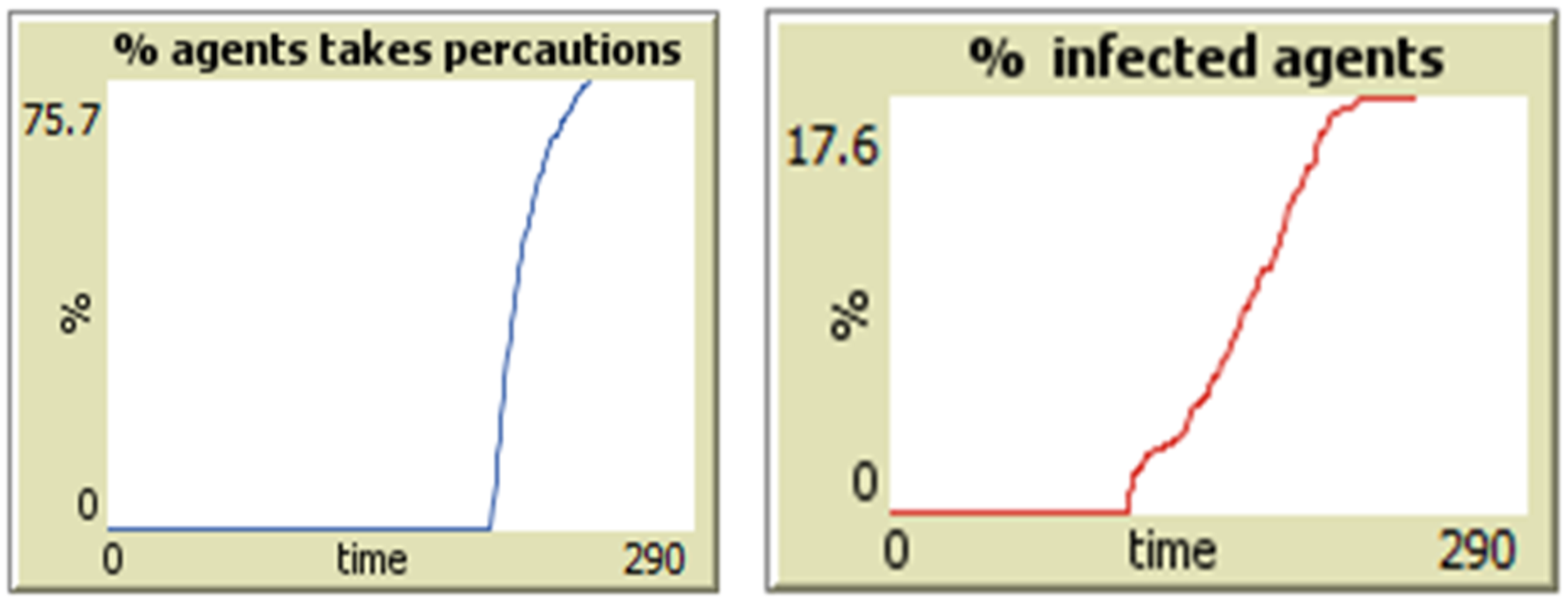}
\caption{Infection growth curve with turtles taking precautions}
\label{fig_precaution_plot_inkscape}
\end{figure}
\begin{figure}[!h]
\centering
\includegraphics[width=\columnwidth]{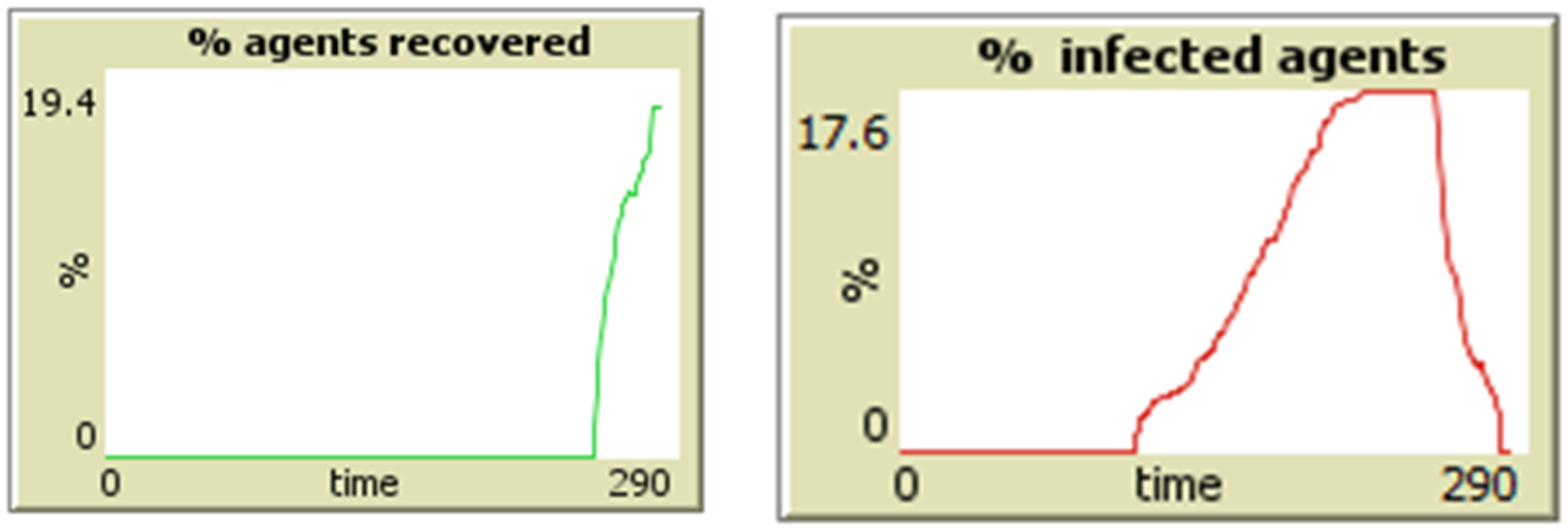}
\caption{Infection growth curve with turtles recovery}
\label{fig_recovery_plot_inkscape}
\end{figure}
Two plots in Fig. \ref{fig_precaution_plot_inkscape} show the behaviour when the infection propagation happens and turtles also take precaution. Infection propagation saturates when turtles start taking precaution.

Fig. \ref{fig_recovery_plot_inkscape} shows two plots which show the behaviour when the recovery process starts and infection among turtles start dropping.
\section{Conclusion and Future Work}\label{section_conclusion}
This paper presented a novel NetLogo model for virus spread addressing features, such as, multi-site mobility, site and route (un)lockdowns, precautions and recovery, which are not available in the existing models. The results presented are only to demonstrate the model functionalities and do not serve any other purpose. This model can be used for education and research purposes. Also, the model may provide a broader framework for more detailed simulations.
More advanced features of NetLogo could reduce the code size. In future, attempts would be made to make some of the hardcoded parameters configurable.

\bibliographystyle{IEEEtran}
\bibliography{IEEEabrv,citation_ref}

\end{document}